# Fansmitter: Acoustic Data Exfiltration from (Speakerless) Air-Gapped Computers

Mordechai Guri, Yosef Solewicz, Andrey Daidakulov, Yuval Elovici
Ben-Gurion University of the Negev
Cyber Security Research Center

gurim@post.bgu.ac.il; yosef.solewicz@gmail.com; daidakul@post.bgu.ac.il; elovici@post.bgu.ac.il

*Abstract*

Because computers may contain or interact with sensitive information, they are often air-gapped and in this way kept isolated and disconnected from the Internet. In recent years the ability of malware to communicate over an air-gap by transmitting sonic and ultrasonic signals from a computer speaker to a nearby receiver has been shown. In order to eliminate such acoustic channels, current best practice recommends the elimination of speakers (internal or external) in secure computers, thereby creating a so-called 'audio-gap'.

In this paper, we present *Fansmitter*, a malware that can acoustically exfiltrate data from air-gapped computers, even when audio hardware and speakers are not present. Our method utilizes the noise emitted from the CPU and chassis fans which are present in virtually every computer today. We show that a software can regulate the internal fans' speed in order to control the acoustic waveform emitted from a computer. Binary data can be modulated and transmitted over these audio signals to a remote microphone (e.g., on a nearby mobile phone). We present Fansmitter's design considerations, including acoustic signature analysis, data modulation, and data transmission. We also evaluate the acoustic channel, present our results, and discuss countermeasures. Using our method we successfully transmitted data from air-gapped computer without audio hardware, to a smartphone receiver in the same room. We demonstrated the effective transmission of encryption keys and passwords from a distance of zero to eight meters, with bit rate of up to 900 bits/hour. We show that our method can also be used to leak data from different types of IT equipment, embedded systems, and IoT devices that have no audio hardware, but contain fans of various types and sizes.

## 1. Introduction

Air-Gapped computers are kept physically isolated from the Internet or other less secure networks. Such isolation is often enforced when sensitive or confidential data is involved, in order to reduce the risk of data leakage. Military networks such as the Joint Worldwide Intelligence Communications System (JWICS) [1], as well as networks within financial organizations, critical infrastructure, and commercial industries [2] are known to be air-gapped due to the sensitive data they handle. Despite the high degree of isolation, even air-gapped network have been breached in recent years. While most famous cases are Stuxnet [3] and agent.btz [4], other cases have also been reported [5] [6].

While the breach of such systems has been shown to be feasible in recent years, the exfiltration of data from non-networked computers or those without physical access is still considered a challenging task. Different types of out-of-band covert channels have been proposed over the years, exploring the feasibility of data exfiltration through an air-gap. Electromagnetic methods that exploit electromagnetic radiation from different components of the computer [7] [8] [9] [10] are likely the oldest kind of covert channel researched. Other type of optical [11] and thermal [12] out-of-band channels have also been suggested.

Exfiltration of data using audible and inaudible sound has been proposed and explored by [13] [14] [15]. The existing method suggests transmitting data though the air-gap via high frequency soundwaves emitted from computer speakers. For example, the work in [14] demonstrates a malware (keylogger) that covertly transmits keystroke data through near-ultrasonic audio emitted from laptop speakers. Interestingly, in 2013 security researchers claimed to find BIOS level malware in the wild (dubbed BadBios) which communicates between air-gapped laptops using ultrasonic sound [16].

## 1.1. Speakerless, audioless computers

Acoustic covert channels rely on the presence of audio hardware and a speaker in the transmitter computer. To that end, common practices and security policies prohibit the use of speakers and microphones in a secure computer, in order to create a so-called 'audio-gap' [17] [18]. Motherboard audio support may also be disabled in the BIOS to cope with the accidental attachment of speakers to the line out connectors. Obviously, disabling audio hardware and keeping speakers disconnected from sensitive computers can effectively mitigate the acoustic covert channels presented thus far [19].

In this paper we introduce an acoustic channel which doesn't require a speaker or other audio related hardware to be installed in the infected computer. We show that the noise emitted from a computer's internal CPU and chassis cooling fans can be intentionally controlled by software. For example, a malicious code on a contaminated computer can intentionally regulate the speed of a computer's cooling fans to indirectly control its acoustic waveform. In this way, sensitive data (e.g., encryption keys and passwords) can be modulated and transmitted over the acoustic channel. These signals can then be received by a remote microphone (e.g., via a nearby smartphone), and be decoded and sent to an attacker. Our new method is applicable for a variety of computers and devices equipped with internal fans, including devices such as servers, printers, industrial and legacy systems, and Internet of Things (IoT) devices.

## 2. Related Work

Covert channels have been widely discussed in professional literature [20] [21] [22]. Our work focuses on covert channels that can exfiltrate data from air-gapped computers without requiring network connectivity. Over the years different types of out-of-band covert channels have been proposed, aimed at bridging air-gap isolation. The proposed methods can be categorized into electromagnetic, optic, thermal, and acoustic covert channels.

Electromagnetic emissions are probably the oldest type of covert channels that have been explored academically. Kuhn and Anderson [7], in a pioneer works in this field, discuss hidden data transmission using electromagnetic emissions from a video card. Thiele [23] utilizes the computer monitor to transmit radio signals to a nearby AM receiver. AirHopper [24] exploits video card emissions to bridge the air-gap between isolated computers and nearby mobile phones. GSMem [10], Funthenna [25], and Savat [26] introduce attack scenarios in which attackers can use different sources of electromagnetic radiation from a computer's motherboard, as covert exfiltration channels. Kasmi et al discuss intentional generation of electromagnetic energy in power networks to establish a command and control covert channel with computers [27]. More recently, Matyunin et al propose using the magnetic field sensors of mobile devices as a covert channel [28]. Loughry and Umphress [11] discuss information leakage from keyboard LEDs through optical emanations. Shamir et al demonstrated how to establish a covert channel with a malware through the air-gap using remote laser and scanners [29]. BitWhisper [30] demonstrates a covert communication channel between computers via exchange of so-called 'thermal-pings.' Characterizing and measuring out-of-band covert channels has been discussed in [31] [32].

## 2.1. Acoustical methods

Madhavapeddy et al [15] discuss 'audio networking,' which allows data transmission between a pair of desktop computers, using 10 dollar speakers and a microphone. In 2013, Hanspach and Goetz [33] extend a method for near-ultrasonic covert networking between air-gapped computers using speakers and microphones. They create a mesh network and use it to implement an air-gapped key-logger to demonstrate the covert channel. The concept of communicating over inaudible sounds has been comprehensively examined by Lee et al [13], and has also been extended for different scenarios using laptops and smartphones [34] . Table 1 summarizes the different types of covert channels for air-gapped computers, including our Fansmitter method.

Table 1. Different types of covert channels for air-gapped computers.

| Method | Examples | Transmitter | Comments |
|---|---|---|---|
| **Electromagnetic** | [7] [8] [10] [25] [26] [24] | Internal computer components | |
| **Optical** | [11] [29] | LEDs | |
| **Thermal** | [12] | CPU, GPU | |
| **Acoustic** | [14] [13] [34] [35] | Internal or external computer Speaker | Audio hardware and speakers are required. |
| **Fansmitter (this paper)** | | Computer fans | No need for audio hardware or speakers. |

As can be seen, existing acoustic methods require the installation of an external or internal speaker in the transmitting computer. This is considered a restrictive demand, because in many cases, speakers are forbidden on air-gapped computers based on regulations and security practices [17]. On the other hand, with our method the transmitting computer does not need to be equipped with audio hardware or an internal or external speaker.

## 3. Attack Model

Although Fansmitter makes a general contribution to the field of acoustic covert channels, like other works which present out-of-band covert channels [14] [13] [24], we propose Fansmitter as a method of exfiltrating information from air-gapped computers. Similar to other covert communication channels, the adversarial attack model consists of a *transmitter* and a *receiver*. Typically in our scenarios, the transmitter is an ordinary desktop computer, and the receiver is a nearby mobile phone (Figure 1). In a preliminary stage, the transmitter and receiver would have been compromised by the attacker. Infecting a highly secure network can be accomplished, as demonstrated by the attacks involving Stuxnet [36], Agent.Btz [37], and others [38] [39] [40]. In our case, the infected computer must be equipped with an internal CPU or chassis fan, a situation which exists in virtually every computer today. Infecting a mobile phone is a much less challenging task and can be done via many different attack vectors, using emails, SMS/MMS, malicious apps, and so on [41] [42] [43] [44].

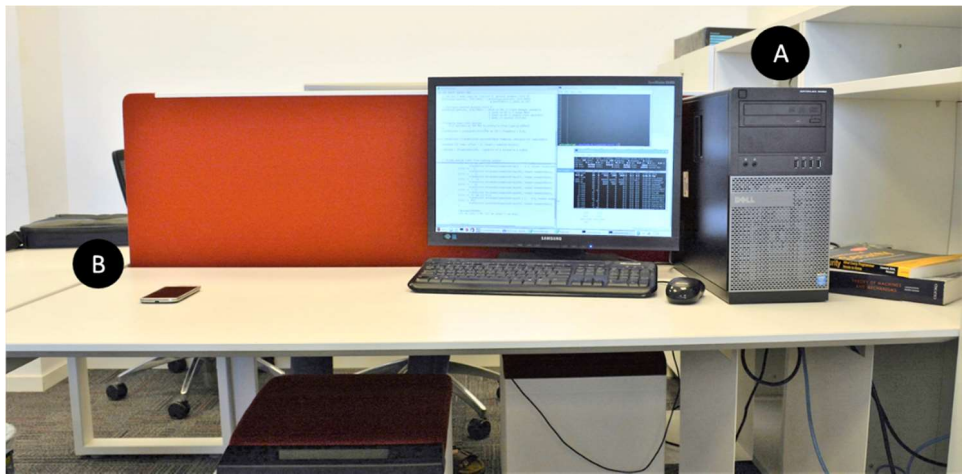

Figure 1. A typical exfiltration scenario. A compromised computer (A) - without speakers, and with audio hardware disabled - transmits sensitive information via acoustic signals. This information is received and decoded by a nearby mobile phone (B)

Then the compromised computer gathers sensitive data (e.g., encryption keys) and modulates and transmits them using the acoustic soundwaves emitted from the computer's internal fans. A nearby mobile phone receiver (also compromised) equipped with a microphone detects and receives the transmission, demodulates and decodes the data, and transfers it to the attacker via mobile data, SMS, or Wi-Fi. Note that in this paper, we demonstrate the attack model using a mobile phone receiver, a device which is commonly located in the vicinity of a computer. Other types of receivers are computers with microphone, laptops, and so on.

## 4. Technical Background

Different computer components including the CPU, RAM and GPU and others produce heat during normal use. These components must be kept within a specified temperature range in order to prevent overheating, malfunction, and damage. The fans are attached to other computer components and are aimed at accelerating the cooling these components by increasing air flow.

Desktop computers are typically equipped with three to four types of fans (power supply unit, chassis, and CPU fans, as well as the optional graphics card fan). Figure 2 illustrates the location of the various fans in a computer.

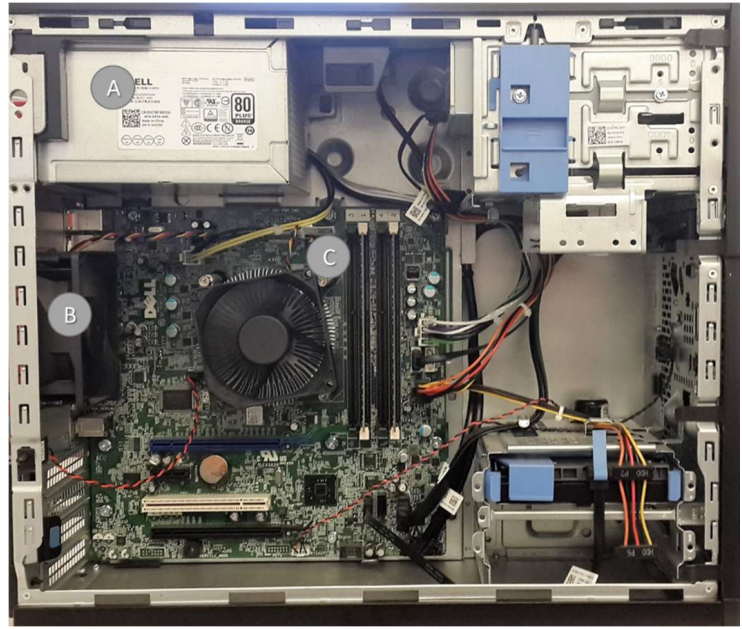

Figure 2. Fan locations within a standard workstation (additional details below)

**The PSU (power supply unit) fan (Figure 2, A).** This fan is integrated into the PSU at its rear part. It is used as an exhaust fan to expel warm air from the PSU which produces heat. This type of fan is managed by an internal controller and usually cannot be monitored, controlled, or regulated by software.

**The Chassis fan (Figure 2, B).** This fan is installed on the side or at the rear of the computer case. It usually draws in cold air from outside the computer and expels it through the top or rear of the computer.

**The CPU fan (Figure 2, C).** This fan is mounted on top of the CPU socket. It cools the CPU's heatsink.

**The GPU (computer graphics card) fan.** Due to the high power consumption and heat emission of modern graphics cards, dedicated cooling fans are installed. Like CPU fans, they are mounted to the heatsink of the GPU.

In addition, there are other fans which are less common, such as: HD, PCI, optical driver, and memory slot fans. These fans can be seen on servers or legacy equipment.

In this paper we focus on CPU and chassis fans. These fans are guaranteed to be present in every computer, making Fansmitter a threat to virtually every computer. The PSU fan has been omitted from our discussions due to its complete separation from the motherboard.

### 4.1. Fan control

Chassis and processor fans use either three or four wire (female) connectors, which are connected to three or four pin headers on the motherboard. Table 2 lists the four wires and their functionality. Note that most modern motherboards are shipped with four wire CPU and chassis fans.

Table 2. Connectors of four wire fans and their functionality

| Pin | Symbol | Function |
|---|---|---|
| **1 (red)** | GROUND | Ground wire |
| **2 (black)** | 12 V | 12 V fan powering |
| **3 (black)** | FAN_TACH | Output signal reporting the fan speed |
| **4 (yellow)** | FAN_CONTROL | Input signal, allows adjustment of the fan speed |

As can be seen, the third wire is used as an output signal, which continuously reports the fan speed to the motherboard. The fourth wire is used as an input signal, allowing setting and adjustment of the fan speed via a pulse-width modulation signal (PWM) [45]. In older three wire fans, the FAN_CONTROL wire is not present, and hence fan speed is not controllable.

The speed of CPU and chassis fans can be managed automatically or manually. With automatic management, the fan speed is regulated by a controller in the motherboard. The controller increases and decreases the fan speed according to the current temperature. Manual management can override the current fan speed and set it using the fan control input signal. Setting the fan speed manually is usually done via the BIOS interface or from within the OS, if an appropriate driver to access the appropriate bus has been installed. There are several open-source and propriety programs aimed at controlling and monitoring fan speed in different motherboards for Windows, Linux, and Mac OSs [46] [47] [48].

## 5. Fan Rotation and Acoustic Signal

Computer fan rotation, measured in revolutions per minute (RPM) units, emits acoustic noise at various frequencies and strength. Typical computer fan speeds range from a few hundred revolutions per minute to a few thousand RPMs. The noise is caused by the movement of the fan blades, each of which pushes air in its path, and together their movements create a compression wave with some amount of rarefaction. The noise level depends on the air flow, mechanics, and vibrations which are mainly determined by the location, size, number of blades and the current RPM of the rotating fan. Because the location, size, and number of blades of a fan are fixed, the current RPM is the main factor which determines the noise level. It is known that fan noise (in dB) increases with the fifth power of the fan rotation speed [49], hence changes in the fan's RPM cause an immediate change to the noise emitted.

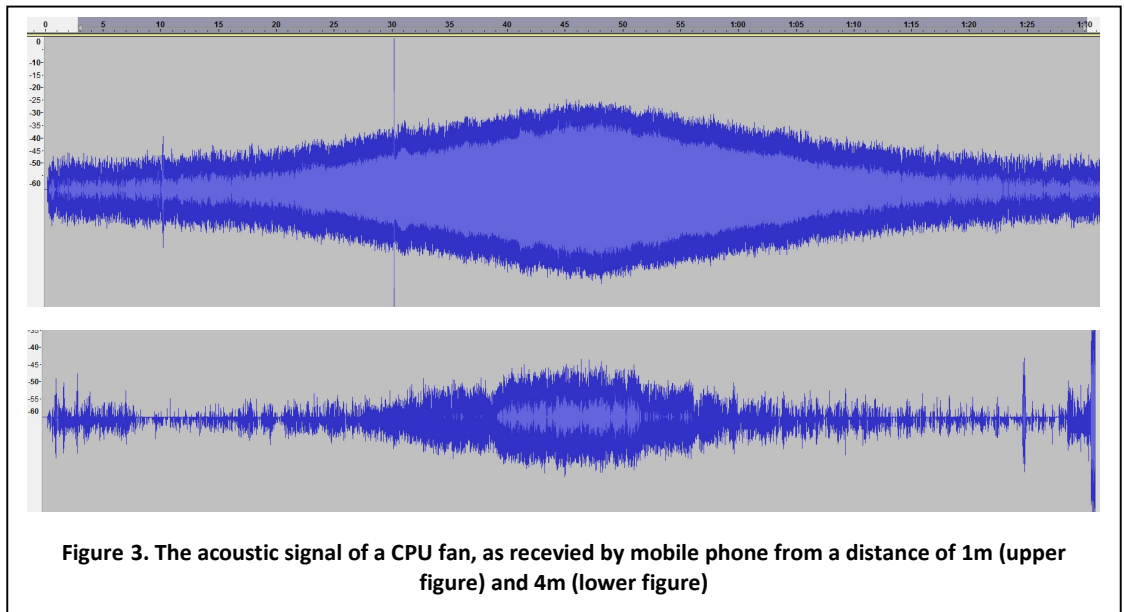

**Figure 3. The acoustic signal of a CPU fan, as recevied by mobile phone from a distance of 1m (upper figure) and 4m (lower figure)**

Figure 3 displays the waveforms produced by accelerating a computer CPU fan from 1000 RPM to 4500 RPM and back. The sounds produced were recorded by a mobile phone placed one meter (upper figure) and four meters (lower figure) away from the computer fan.

Note that although the acoustic waveform may differ between various types of fans, the basic phenomenon observed in which an increased rate of fan rotation is associated with increased acoustic energy in the wave produced is common to all types of rotating fans [49]. This permits us to extend the concept presented in this paper to other types of fans in a variety of devices.

### 5.1. Blade pass frequency

Blade pass frequency (BPF) is the main acoustic tone generated by a running fan. More specifically, it is the tone frequency produced when the blades of the fan rotate past the turning vanes. The overall noise spectrum pattern of a fan's noise can be broken down into several components for the purpose of analysis. The BPF itself, while not being a perfect sinusoidal wave, generates harmonics at multiples of its frequency. In addition, unsteady flows contribute with both noises spread around the BPF and high frequency broadband noise.

BPF (in $Hz$) is calculated by multiplying the number of blades ($n$), with the rotating speed ($R$) in revolutions per minute (RPM), so $BPF = n * R/60$.

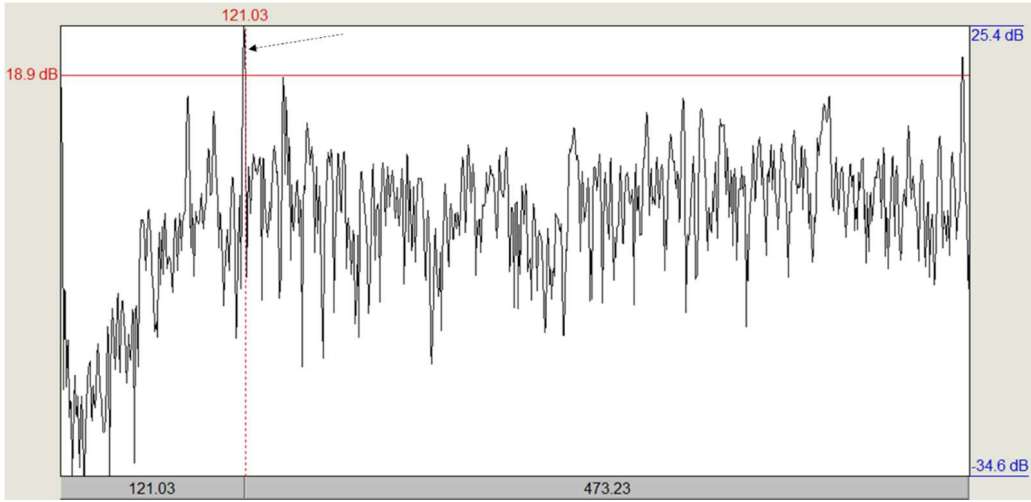

**Figure 4. Typical BPF plot (R ~ 1000 RPM)**

Figure 4 illustrates a typical fan spectrum at about 1000 RPM. The BPF is indicated in red at around $120\ Hz$, as expected for a seven blade fan.

### 5.2. Signal processing

In order to analyze the waveforms in the frequency domain we proceeded as follows. First, for simplicity, we downsampled the signals from the original $44.1\ kHz$ sampling frequency recording to $2\ kHz$. The reduced sample rate is sufficient for the analysis of signals below $1\ kHz$ (half of the sample rate as is prescribed by the Nyquist theorem), which is still above the expected acoustic tone corresponding to the rotation levels used, as explained above. Second, we bandpass filtered the signals around the expected BPF, namely from $400$ to $600\ Hz$.

The BPF enables us to calculate the expected tone generated by the fan. It also allows us to use a frequency based modulation scheme. For example, since a typical chassis fan has seven blades ($n = 7$), the expected BPF values for the $R$ ranges used in our experiments can be precalculated (see Table 3).

**Table 3. Expected BPF values for seven blade fans**

| R (RPM)   | BPF (Hz) |
|-----------|----------|
| 1000-1600 | 116-187  |
| 1600-3000 | 187-350  |
| 2000-2500 | 233-292  |
| 4000-4500 | 466-525  |

### 5.3. Channel stealth

Based on the BPF formula, a typical computer fan emits noise in the range of $100$ to $600\ Hz$, a frequency range which can be detected by the human ear. The fan's emitted noise is noticeable at a high frequency range, because the noise level increases with the fifth power of the fan rotation speed [49]. By definition covert channels must work stealthily in order to avoid detection. There are three strategies which can be implemented to enable the channel to operate in stealth mode.

1. **Flexibility in terms of the time of the attack.** The attacker may choose to start the transmissions at times when users are not working on the computer, or even present in the room (e.g., during evening hours). Note that this strategy works with some attack models [11] [29]. This relevant also in cases in which the attacker has a persistent microphone in the room, such as other computers equipped with a microphone.

2. **The use of low frequencies.** Modulating the data over lower frequencies (e.g., $140 - 170\ Hz$) appears to be less noticeable by users, especially when the receiver is located at short distance from the transmitting computer. In some scenarios, an attacker may choose to only use low frequencies for the transmissions.

3. **The use of close frequencies.** Modulating the data over change of close frequencies (e.g., less than a $100\ Hz$ difference between 0 and 1) is also less noticeable by a user, as it blends in and appears as natural background environmental noise.

# 6. Data Modulation

To encode digital information over the fan's acoustic signal, the data must be *modulated*. In our case we modulate the binary data over a *soundwave* carrier: the fan sound waveform. We use two types of modulation schemes: (1) Amplitude Shift Keying (ASK), and (2) Frequency Shift Keying (FSK). FSK modulation is faster and more resilient to environmental noises than ASK, while ASK is more resilient to the type of fan in use and its blade pass frequency properties than FSK. Generally, we use FSK when the generic type of transmitting fan is known in advance, and ASK is used when the type of fan or its properties are unknown to the attacker. The modulation schemes, in context of our covert channel, are described below.

## 6.1. Amplitude Shift Keying (ASK)

In Amplitude Shift Keying modulation we assign distinct amplitude levels of the carrier wave to represent distinct values of binary data. We use the binary version of ASK (B-ASK), in which two distinct amplitudes, $A_0$ and $A_1$, represent '0' or '1.' We control the carrier amplitudes in a carrier frequency, $F_c$ ($F_a$ to $F_b$), by controlling the fan so that it rotates at two speeds, $R_0$ and $R_1$. Rotation at $R_1$ results in a carrier amplitude of $A_0$ (a logical '0'), while rotation at R1 results in a carrier amplitude of $A_1$ (a logical '1'). Table 4 provides a summary of the parameters of ASK modulation.

Table 4. B-ASK modulation

| RPM | Soundwave Amplitude | Duration | Carrier Freq. | Represent |
|---|---|---|---|---|
| $R_0$ | $A_0$ | $T$ | $F_c$ ($F_a$ to $F_b$) | "0" |
| $R_1$ | $A_1$ | $T$ | $F_c$ ($F_a$ to $F_b$) | "1" |

Figure 5 shows the ASK representation of '101010' modulated by the CPU fan of a desktop computer, as received by a nearby smartphone. In this case, $R_0$=3000 RPM, $R_1$=3500 RPM, and $T$=5 $sec$.

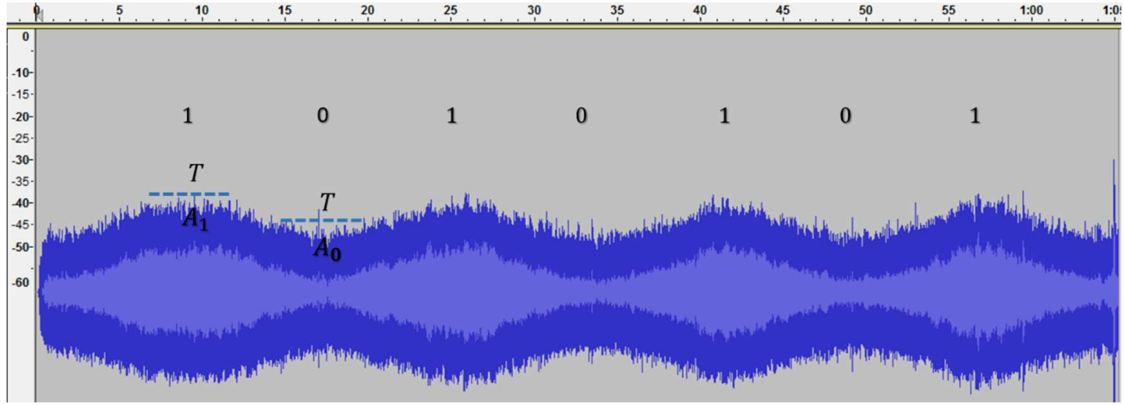

Figure 5. ASK modulation of "101010" over 60 seconds, when $R_0$=3000 RPM and $R_1$=3500 RPM

As noted in [49] $R_0 < R_1$ implies $A_0 < A_1$. Practically, it means that higher RPM rates cause higher noise levels, and thus a higher amplitude carrier wave. Please note that this is a generalization for all CPU fans, and in reality, $R_0$, $R_1$, $A_0$, $A_1$, and $F_c$ may vary for different types of fans.

We only used two RPM rates so as to create two distinct amplitudes to modulate the data. We noticed that sometimes, particularly when the fan is operating at a high speed (e.g., during an heavy workload), it is more effective to assign '0' to high RPM rates and '1' to low RPM rates so that the base amplitude is the higher one.

## 6.2. Frequency Shift Keying (FSK)

In Frequency Shift Keying modulation, we assign distinct frequencies to represent distinct values of binary data. We use the binary version of FSK (B-FSK), in which two distinct frequencies, $F_0$ and $F_1$, represent '0' or '1' arbitrarily. As mentioned, the blade pass frequency is determined by the current RPM, and a change to the RPM implies a change to the current frequency. We maintain the frequency of the carrier by setting the fan to rotate at two speeds, $R_0$ and $R_1$. Rotation at $R_0$ results in a carrier frequency of $F_0$ (a logical '0'), while rotation at $R_1$ results in a carrier amplitude of $F_1$ (a logical '1'). Table 5 provides a summary of the parameters of FSK modulation.

Table 5. B-FSK modulation

| RPM | Carrier Freq. | Duration | Represent |
|---|---|---|---|
| $R_0$ | $F_0$ | $T$ | "0" |
| $R_1$ | $F_1$ | $T$ | "1" |

Figure 6 shows the FSK representation of '101010' modulated by the CPU fan of a desktop computer, as received by a nearby smartphone. In this case, $R_0$=1000 RPM, $R_1$=1600 RPM, and $T$=5 $sec$.

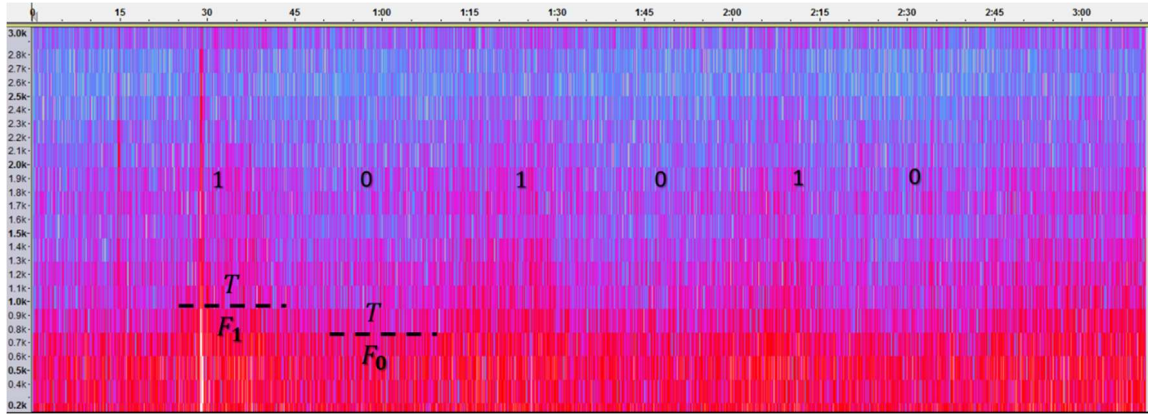

Figure 6. FSK modulation of "101010" over 150 seconds, when $R_0$=1000 RPM and $R_1$=1600 RPM

Based on the BPF formula $R_0 < R_1$ implies $F_0 < F_1$. Practically, it means that higher RPM rates cause a noticeable peak in higher frequencies. As in ASK, this is a generalized form to all CPU fans and $R_0$, $R_1$, $F_0$, $F_1$ and may vary between different types of fans.

## 6.3 Bit framing

As explained in the previous sub-sections, the B-ASK and B-FSK modulation uses two amplitudes and frequencies, respectively, to represent the data. Two issues arise from this:

(1) The amplitudes (in the case of ASK) and frequencies (in the case of FSK) encoding ('0' and '1') depends on the type of fan currently in use and on the exact RPM values used during the modulation. The receiver has no prior information regarding the parameters used by a nearby transmitting computer.
(2) In the case of ASK, the level of the $A_0$ and $A_1$ amplitudes is dependent on the current distance between the transmitting computer and the receiver. This means that if the receiver (e.g., mobile phone) is moving during a transmission, a '1' and '0' may be decoded incorrectly.

To resolve these issues and assist the receiver in dynamically synchronizing with the transmitter parameters, we transmit data in small frames. Each frame consists of a preamble sequence of four bits and a payload of 12 bits (see Table 6).

Table 6. A frame consisting of four bits of preamble, followed by a payload of 12 bits

| Preamble (4 bits) | Payload (12 bits) |
|---|---|
| 1010 | 111010101110 |

The preamble consists of the '1010' sequence and is used by the receiver to determine the amplitude levels or frequency values of a '0' and a '1' in a periodic way. In addition, the preamble header allows the receiver to identify the beginning of a transmission in the area and extract other channel parameters, such as transmission time and carrier frequency.

## 7. Results

We evaluate Fansmitter with a CPU and chassis fan – two fans that exist on virtually every computer today. As noted, due to the acoustic property of computer fans, our method is applicable to other types of fans (e.g., GPU fans) as well. We used a standard desktop computer (DELL OptiPlex 9020) with Intel Core i7-4790 motherboard and Intel Q87 (Lynx Point) chipset, as a transmitter. As a receiver, and to represent an attack model scenario, we used a Samsung Galaxy S4 (I9500) mobile phone with a standard microphone with a sampling rate of $44.1\ kHz$. Our testing environment consisted of a computer lab with ordinary background noise, seven workstations, several network switches, and an active air conditioning system. In practice, the transmission time and bit rate are highly dependent on environmental noise which determines the signal-to-noise ratio ($SNR$). A higher level of environmental noise reduces the bit rate and the channel capacity. The following results based on a distance of one, four, and eight meters between the transmitter and receiver are based the testing environment described above.

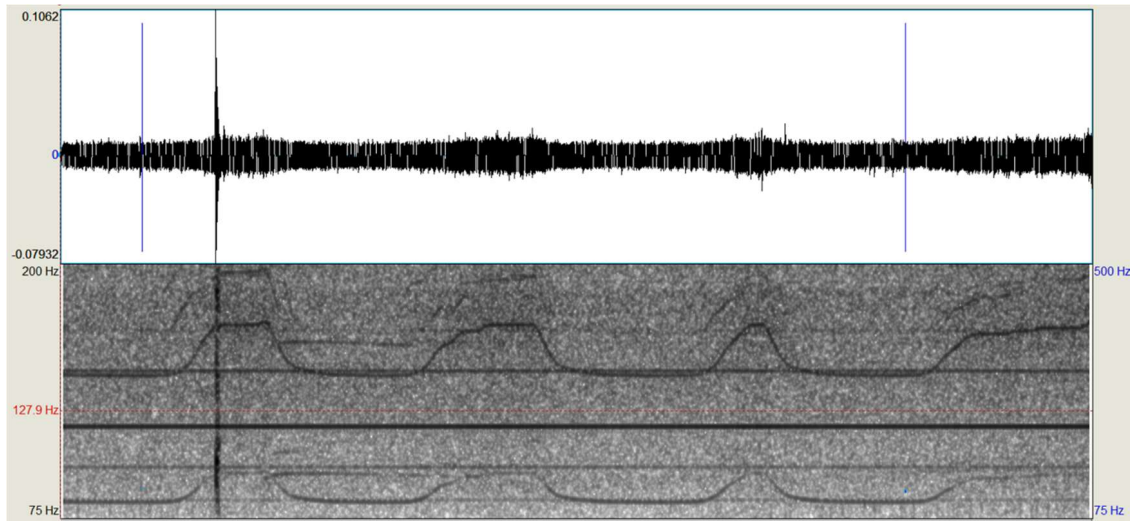

Figure 7. CPU fan audio spectral views (1000-1600 RPM, 1 meter)

Figure 7 depicts a payload (01010101) as received by a mobile phone within a short distance (1 meter) of the transmitting computer. Here we use B-FSK modulation with $R_0 = 1000\ RPM$ and $R_1 = 1600\ RPM$. In this case, $F_0 \approx 145\ Hz$ and $F_1 \approx 170\ Hz$. The total transmission time was 180 seconds with $T = 10\ sec$ which implies an effective bit rate of 3 bits/min. Note that the $F_0$ is not the expected BPF for this specific RPM as shown in Table 3. In fact, the expected $F_0$ is displayed as a continuous frequency peak in the spectrogram. A possible explanation for this may derive from low RPM values which lead to a less predictable noise spectrum.

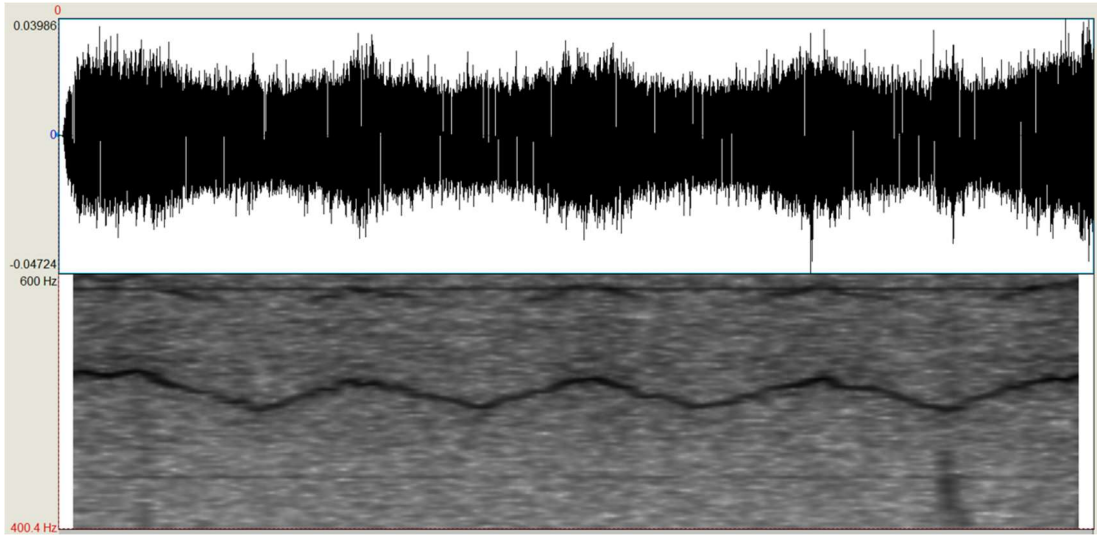

**Figure 8. CPU fan audio spectral views (4000-4250 RPM, 1 meter)**

Figure 8 depicts a payload (01010101) as received by a mobile phone within a short distance of the transmitting computer. Here we use d B-FSK modulation with $R_0 = 4000\ RP$ and $R_1 = 4250\ RPM$. In this case, $F_0 \approx 495\ Hz$ and $F_1 \approx 525\ Hz$. The total transmission time was 32 sec with $T = 0\ sec$ (no delay) which implies an effective bit rate of 15 bits/min (900 bits/hour).

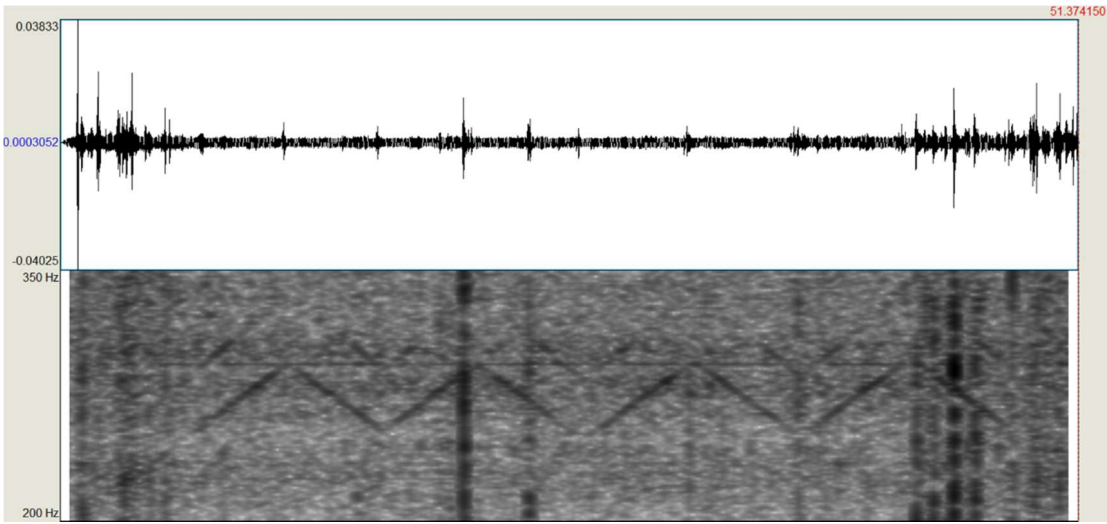

**Figure 9. Fan audio, waveform, and spectral views (2000-2500 RPM, 4 meters)**

Figure 9 depicts a payload (01010101) as received by a mobile phone within a distance of four meters from the transmitting computer. Here we use d B-FSK modulation with $R_0 = 2000\ RPM$ and $R_1 = 2500\ RPM$. In this case, $F_0 \approx 261\ Hz$ and $F_1 \approx 294\ Hz$. The total transmission time was 52 seconds with $T = 0\ sec$ (no delay) which implies an effective bit rate of 10 bits/min (600 bits/hour).

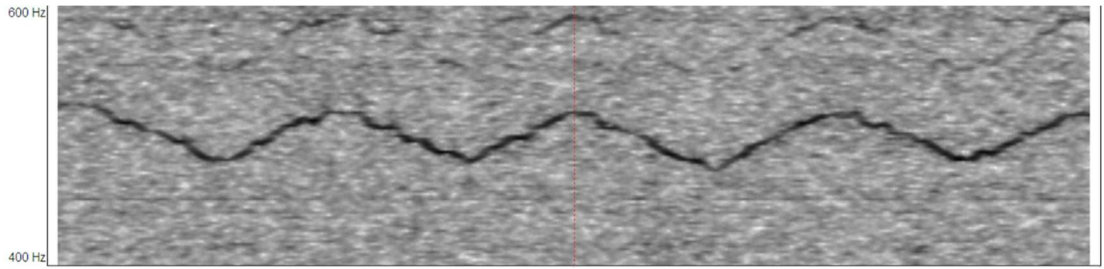

**Figure 10**. Fan audio, waveform, and spectral views (4100-4500 RPM, 8 meters)

Figure 10 depicts a payload (10101010) as received by a mobile phone within a distance of four meters from the transmitting computer. Here we use d B-FSK modulation with $R_0 = 4100$ and $R_1 = 4500\ RPM$. In this case, $F_0 \approx 480 Hz$ and $F_1 \approx 527\ Hz$. The total transmission time was 48 sec with $T = 0\ sec$ (no delay) which implies an effective bit rate of 10 bits/min (600 bits/hour).

## 7.1. Bit rate

The effective bit rate is determined by the time it takes the fan to transit between two level of speeds, and bit transmission time. Let $T$ be the duration (in seconds) of '0' and '1' transmissions. Let $TR_{0-1}$ be the time (in seconds) it takes the fan rotor to transfer from speed $R_0$ to $R_1$ and $TR_{1-0}$ be the time it takes the fan rotor to transfer from $R_1$ to $R_0$. To simplify the calculations we assume that $TR = TR_{0-1} = TR_{1-0}$. Then, the time it takes to transmit $n$ bits is $t \approx n * (TR + T)\ \frac{Bits}{Second}$.

## 8. Countermeasures

There are three main categories of countermeasures: procedural, software based, and physical or hardware based.

In procedural countermeasures the zones [50] approach may be used. In this approach sensitive computers are kept in restricted areas in which mobile phones, microphones, and electronic equipment are banned. The zones approach (also referred to as the 'black-red separation' approach) is discussed in [51] as a means of handling various types of acoustic, electromagnetic and optical threats. However, zones are not always possible, due to practical reasons and space limitations. Software based countermeasures include the use of endpoint protection to detect malicious activities on the computer, and interfering with the fan control API and bus access. Such countermeasures have been shown to be porous [52] as an attacker can use rootkit techniques and other techniques to evade detection. Fan speed regulation by software (e.g., device driver) can regulate fan speed and therefore noise emission, hence preventing covert data modulation. Hardware based and physical countermeasures may include noise detectors which monitor soundwave at specific or range of frequencies. Such products exist [53] but are prone to false alerts from environmental noise. Jamming the fan signal by the generation of background noise [54] is also possible but not applicable in some environments, particularly in quiet settings. Physical isolation in which the computer chassis is built with special noise-blocker coverage is also an option, but this is costly and not practical on large-scale. Finally, replacing the fans with specialized quiet fans can limit the noise level, but even this does not completely prevent the emission of noise [49]. Moving to different types of cooling systems, such as water cooling or refrigeration can prevent acoustical emission, but this is also less practical for wide deployment. Table 7 summarizes the listed countermeasures, along with challenges of each.

Table 7. Procedural, software, and hardware countermeasure

| Method | Type | Challenges |
|---|---|---|
| **'Zones' separation** | Procedural | Space limitations |
| **AV / monitoring** | Software | Can be bypassed by rootkits or evasion techniques |
| **Fan regulation** | Software | Can be bypassed by low-level malware |
| **Noise detection** | Hardware | False positives and false alarms |
| **Signal jamming** | Hardware | Generate background noises |
| **Fan replacement, Water cooling, etc.** | Hardware | Financial limitations |
| **Chassis isolation** | Physical | Financial and space limitations |

# 9. Conclusion

Past research has demonstrated that malware can exfiltrate information though an air-gap by transmitting audio signals from the internal or external speakers of desktop computers. In this paper, we propose a method that uses the noise emitted from computer fans as a medium for acoustic data exfiltration. We show that the acoustic waveform emitted from CPU and chassis fans of a desktop computer can be regulated and controlled. We also show that a malware can modulate information over the waveform and transmit it to a nearby receiver. We designed a simple amplitude and frequency based modulation scheme and tested it in an actual workplace setting with ordinary noise levels. Our tests shows that information be received by a smartphone (or a similar device with a microphone) from a distance of zero to eight meters and at a bit rate of up to 15 bits per minute. Using Fansmitter attackers can successfully exfiltrated passwords and encryption keys from a speakerless air-gapped computer to a mobile phone in the same room from various distances. Beyond desktop computers, our method is applicable to other kinds of audioless devices, equipped with cooling fans (various types and sizes of fans) such as printers, control systems, embedded devices, IoT devices, and more.